\documentclass[a4paper]{JHEP} 

\usepackage{epsfig}

\newcommand\fverb{\setbox\pippobox=\hbox\bgroup\verb}
\newcommand\fverbdo{\egroup\medskip\noindent%
                        \fbox{\unhbox\pippobox}\ }
\newcommand\fverbit{\egroup\item[\fbox{\unhbox\pippobox}]}
\newbox\pippobox

\newcommand{\beq}{\begin{equation}}
\newcommand{\eeq}{\end{equation}}
\newcommand{\ba}{\begin{array}}
\newcommand{\ea}{\end{array}} 
\newcommand{\beqa}{\begin{eqnarray}}
\newcommand{\eeqa}{\end{eqnarray}}

\newcommand{\cL}{{\cal L}}
\newcommand{\cH}{{\cal H}}

\newcommand{\cO}{{\cal O}}
\newcommand{\cM}{{\cal M}}
\newcommand{\da}{^\dagger}

\newcommand{\lsim}{\stackrel{<}{_\sim}}
\newcommand{\gsim}{\stackrel{>}{_\sim}}
\newcommand{\Ko}{{K}^0}
\newcommand{\Kob}{\bar{K}^0}

\newcommand{\tq}{{\tilde q}}
\newcommand{\tl}{{\tilde l}}

\newcommand{\tu}{{\tilde u}}
\newcommand{\deu}{(\delta^{U}_{LR})}
\newcommand{\deull}{(\delta^{U}_{LL})}

\newcommand{\PL}[3]{{Phys. Lett.}       {\bf #1} {(19#2)} {#3}}
\newcommand{\PRe}[3]{{Phys. Rep.}       {\bf #1} {(19#2)} {#3}}
\newcommand{\PRL}[3]{{Phys. Rev. Lett.} {\bf #1} {(19#2)} {#3}}
\newcommand{\PR}[3]{{Phys. Rev.}        {\bf #1} {(19#2)} {#3}}
\newcommand{\NP}[3]{{Nucl. Phys.}       {\bf #1} {(19#2)} {#3}}
\newcommand{\ZP}[3]{{Z.  Phys.}         {\bf #1} {(19#2)} {#3}}

\newcommand{\PTP}[3]{{Prog. Theor. Phys.} {\bf #1} {(19#2)} {#3}}

\title{Supersymmetric contributions to rare kaon decays:
               beyond the single mass--insertion approximation.}

\author{G. Colangelo\thanks{Address after September 1, 1998:
Institut f\"ur Theoretische Physik der Universit\"at Z\"urich,
Winterthurerstr. 190, CH--8057 Z\"urich--Irchel.} and G. Isidori \\ INFN,
Laboratori Nazionali di Frascati, \\ I--00044 Frascati, Italy \\ E-mail:
\email{colangelo@lnf.infn.it, isidori@lnf.infn.it}}

\preprint{LNF-98/026(P)}      

\abstract{
We analyze the contributions to rare kaon decays mediated by 
flavor--changing $Z$--penguin diagrams in a generic 
low--energy supersymmetric extension of the Standard Model.
In order to  perform a model--independent analysis we expand the squark 
mass matrices around the diagonal, following the so
called mass--insertion approximation. We argue that in the present case it is 
necessary to go up to the second order in this expansion to take into 
account all possible large effects. The current bounds on such
second--order term, which was neglected in previous analyses,
are discussed in detail and the corresponding upper bounds for the
rare kaon decay rates are derived. As a result, we show that 
supersymmetric effects could lead to large enhancements of 
$K\to\pi\nu{\bar \nu}$ and $K_L \to\pi^0 e^+ e^-$ branching ratios.}


\begin{document}

\section{Introduction}   

Flavor--changing neutral--current (FCNC) processes provide
a powerful tool for indirect searches of New Physics. This 
is particularly true in the framework of 
low energy supersymmetry 
\cite{Susy}, which represents one of the most interesting extensions 
of the Standard Model (SM). 
The large number of new 
particles carrying flavor quantum numbers,
present in this context, would naturally lead to sizable
effects in FCNC transitions \cite{SusyFCNC,HKR}. 

At the one--loop level, supersymmetric contributions to FCNC 
amplitudes can be classified into three groups, 
according to the virtual particles inside the loop:
i) Higgs/W--quarks, ii) gluino--squarks
and iii) chargino/neutralino--squarks. The first group  
contains the SM contributions as a particular subgroup,
whereas ii) and iii) represent genuine supersymmetric effects.
Among them, gluino--squark transitions have been widely discussed 
in the literature \cite{HKR,GGMS,MPR,Bagger} and are expected to produce the 
dominant non--SM  effect in $\Delta F=2$ processes. 
This is confirmed by the analysis of $\Ko-\Kob$ mixing with the 
inclusion of gluino--squark contributions, which provides severe 
constraints on supersymmetric models \cite{HKR,GGMS,MPR,Bagger}. 
The effect of chargino/neutralino--squark diagrams is usually neglected 
in the analysis of such processes.

A different situation occurs in $\Delta F=1$ transitions
mediated by $Z$--penguin diagrams, which are particularly 
relevant to rare kaon decays, like $K\to\pi\nu{\bar \nu}$.
As recently discussed in \cite{NW,BRS}, the dominant supersymmetric 
contribution to these processes is given by chargino--up--squarks diagrams. 
This is because the $Z{\bar q}_i q_j$ effective vertex is necessarily 
proportional to $SU(2)_L$--breaking couplings that, in supersymmetric models, 
are provided by $q_L-q_R$, $\tq_L-\tq_R$ and 
wino--higgsino mixing. Since the $\tq_L-\tq_R$
mixing in the down sector is suppressed by the small down--type
Yukawa couplings, the effect of gluino and neutralino 
diagrams is necessarily small. On the other hand, the large Yukawa 
coupling of the top leads to potentially large effects in diagrams 
involving up--type quarks or squarks. Indeed the $(d,s)_L - t_R$
mixing, already present in the SM, is responsible for the 
$m^2_t$ enhancement of the  Higgs/W--quark contribution
to the $Z{\bar s}d$ effective vertex. Analogously, 
it is natural to expect a large effect due to 
the $({\tilde d},{\tilde s})_L-{\tilde t}_R$ mixing
in diagrams involving charginos and up--type squarks. 

This effect has been already noted by Buras, Romanino and Silvestrini in
the calculation of the supersymmetric contributions to $K\to\pi\nu{\bar
\nu}$ \cite{BRS}.  However, this calculation has been performed in the
{\em single} mass--insertion approximation, where only terms with at most one
off--diagonal element of the squark mass matrix are considered.  We believe
that this approximation is not sufficient to fully account for possible
large effects in the present case. Indeed, in order to provide the
necessary $SU(2)_L$ breaking, at least two mass--mixing terms are
necessary, either from the squark sector (${\tilde u}_L-{\tilde u}_R$) or
from the chargino sector (wino--higgsino). Since both these couplings
vanish as $\cO(m_W/M_S)$ in the limit of a heavy supersymmetry--breaking
scale ($M_S$), we consider more appropriate to expand in both of them up to
the second order. One could argue that wino--higgsino mixing is not
suppressed by the off--diagonal flavor structure. However, the hierarchy of
the Yukawa couplings implies that terms with a single wino--higgsino mixing
always appear together with suppressed CKM factors.  As a result, it is
reasonable to expect that terms with a double LR mass insertion and without
any CKM suppression are at least of the same order as those generated by a
single LR mass insertion together with wino--higgsino mixing.

In the present paper we present a complete discussion of the 
supersymmetric contributions to the $Z{\bar s}d$ amplitude,
beyond the single mass--insertion approximation. We find that 
the contribution generated by a double LR mass insertion in the 
up--squark sector, which was neglected in previous analyses,
yields a potentially large effect. Employing the notation of \cite{BRS},
we can naively say that in this case the $\lambda_t m_t^2/m_W^2$ 
factor of the SM amplitude gets replaced by
$(M^2_U)_{s_L t_R} (M^2_U)_{t_R d_L}/M_S^4$.
Interestingly, this kind of mechanism is only weakly constrained 
by $\Ko-\Kob$ mixing and can provide a sizable enhancement
(up to two orders of magnitude) to rare decay widths.

The paper is organized as follows. In section 2 we discuss
the supersymmetric contributions to the $Z{\bar s}d$ amplitude, 
with particular attention to the hierarchy of the various terms.
The role of box diagrams in $\Delta F=1$ transitions is also briefly
analyzed. In section 3 we discuss theoretical and phenomenological
bounds on the up--type LR couplings. In section 4 we analyze 
the possible enhancements of rare kaon decays rates
driven by these supersymmetric effects. 
The results are summarized in the conclusions.

\section{The $Z{\bar s}d$ effective vertex}
\label{sect:Zsd}

The amplitude we are interested in here is the one--loop FC effective
coupling of the $Z$ boson to down--type quarks, in the limit of vanishing
external masses and momenta. As already emphasized in \cite{NW,BRS}, 
the $SU(2)_L \times U(1)_Y  \to U(1)_{e.m.}$ gauge structure implies 
that this coupling proceeds through symmetry--breaking terms and involves 
only left--handed quarks. Thus it can be generally described by
introducing the effective Lagrangian
\beq
\cL_{FC}^Z =  {G_F\over \sqrt{2}} \frac{e}{2\pi^2} M_Z^2
\frac{\cos\Theta_W}{\sin\Theta_W}~
W_{ds}~ Z_\mu {\bar s} \gamma^\mu (1-\gamma_5) d~+~\mbox{h.c.}
\label{eq:Lzsd}
\eeq
where $W_{ds}$ is a complex dimensionless coupling.

In our conventions, the SM contribution of top--quark penguin diagrams,
evaluated in the 't Hooft--Feynman gauge, leads to 
\beq
W_{ds}^{SM}   = \lambda_t C(x_{tW})~,
\label{eq:SM}
\eeq
where $\lambda_t=V_{ts}^* V_{td}$, $V_{ij}$ are the CKM matrix elements
\cite{CKM} and  $x_{tW}=m_t^2/m_W^2$. The loop function $C(x)$, originally 
computed in \cite{ILim}, can be found in the appendix. 
We recall that $C(x) \to x/8$ for large $x$.  

In the minimal supersymmetric extension of the SM, 
which requires two Higgs doublets,
the contribution of penguin diagrams with the exchange of 
charged Higgs and top--quark is aligned with the SM one (i.e. is
proportional to $\lambda_t$). Denoting
as usual by $\tan\beta$ the ratio of the two Higgs vacuum 
expectation values \cite{Susy}, we find
\beq
W_{ds}^{H}   = \lambda_t \frac{m_H^2}{m_W^2 \tan^2\beta}~ H(x_{tH})~,
\label{eq:Higgs}
\eeq 
where now $x_{tH}=m_t^2/m_{H^\pm}^2$. Similarly to the SM case, also
$H(x)\to x/8$ for large $x$ (the full expression of $H(x)$ is given in the
appendix).  The sum of (\ref{eq:SM}) and (\ref{eq:Higgs}) complete the
first class of contributions outlined in the introduction, namely the
Higgs/W--quark diagrams. To analyze the genuine supersymmetric effects, and
particularly those generated by chargino--squark exchange, we first need to
discuss shortly the structure of the supersymmetric mass matrices.

In the basis of the electroweak eigenstates, wino and higgsino, the
chargino mass matrix is given by 
\beq 
M_\chi = \left( \begin{array}{cc} M_2
& \sqrt{2} m_W \sin\beta \\ \sqrt{2} m_W \cos\beta & \mu \end{array}
\right)~, 
\eeq 
where the index 1 of both rows and columns refers to the
wino state. Following the standard notation \cite{Susy}, here $\mu$ denotes
the Higgs quadratic coupling and $M_2$ the soft supersym\-me\-try--breaking
wino mass.  To define the mass eigenstates we introduce the unitary
matrices ${\hat U}$ and ${\hat V}$ which diagonalize $M_\chi$ \beq {\hat
U}^* M_\chi {\hat V}\da = \mbox{diag}(M_{\chi_1},M_{\chi_2})~.  \eeq As can
be noticed, the off--diagonal entries of $M_\chi$ are $\cO(m_W)$, whereas
$M_2$ is $\cO(M_S)$. In the limit where $m_W/M_2$ is a small parameter we
can perform a perturbative diagonalization of $M_\chi$ around its diagonal
elements, or, correspondingly, an expansion of ${\hat U}$ and ${\hat V}$
around the identity matrix.

In the squark sector we have $6\times 6$ matrices which mix the 
three families of left--handed and right--handed squarks. 
A convenient basis
for our calculation is the basis where the $d_L^i-\tu_L^j-\chi_n$ 
coupling is flavor diagonal and the $d_L^i-\tu_R^j-\chi_n$ 
one is ruled by the CKM matrix (see \cite{BRS} for a more detailed
description). In this case, the up--squark mass matrix 
is given by the Hermitian matrix  
\beq
\cM^2_U = \left( \begin{array}{cc} (M^2_U)_{d_L d_L} & (M^2_U)_{d_L u_R} \\
(M^2_U)_{u_R d_L} & (M^2_U)_{u_R u_R} \end{array} \right)
\eeq
where the subscript $d_L$ (which runs over three values)
indicates the combination of left--handed up--type squarks 
which appear in the diagonal couplings 
$d_L-\tu^{(d)}_L-\chi_n$. On the other hand, the index $u_R$
denotes the combination of right--handed up--type squarks which 
appear in the $d_L^i-\tu_R^j-\chi_n$ vertices ruled by
the CKM matrix--element $V_{ij}$. 
Since $\cM^2_U$ is Hermitian we need to introduce
only one unitary matrix, ${\hat H}$, to diagonalize it 
\beq
{\hat H}\cM^2_U {\hat H}\da = \mbox{diag}\left(M_{\tu_1},M_{\tu_2},\ldots,
M_{\tu_6}\right)~. 
\eeq
Also for $\cM^2_U$ the off--diagonal elements are expected to be small
and the perturbative diagonalization is well justified \cite{HKR}. 

We are now ready to evaluate the contribution of the chargino--squark
penguin diagrams in Fig.~\ref{fig1}. The full result before any mass
expansion is quite simple and is given by\footnote{~The sum over the
repeated indices $i$ and $j$ (running from 1 to 2), $l$ and $k$ (running
from 1 to 6), and $q_L$ (running over the three values $d_L$, $s_L$ and
$b_L$ ) is understood.}  
\beq W_{ds}^{\chi} = {1\over 8} A^d_{jl}{\bar A}^s_{ik}
F_{jilk}~,
\label{eq:Wchi}
\eeq
where
\beqa
A^d_{jl} &=& {\hat H}_{l d_L}{\hat V}\da_{1j} 
            - g_t V_{td}{\hat H}_{l t_R}{\hat V}\da_{2j}~, \\
{\bar A}^s_{ik} &=& {\hat H}\da_{s_L k}{\hat V}_{i1} 
            - g_t V_{ts}^*{\hat H}\da_{t_R k}{\hat V}_{i2}~, \\
F_{jilk} &=& {\hat V}_{j1}{\hat V}_{1i}\da~ \delta_{lk}~ k(x_{ik},x_{jk}) 
             - 2 {\hat U}_{i1} {\hat U}\da_{1j}~
\delta_{lk}~ \sqrt{x_{ik}x_{jk}} j(x_{ik},x_{jk}) \nonumber \\
         &&  -\delta_{ij}~ {\hat H}_{k q_L} {\hat H}\da_{q_L l}~ k(x_{ik},x_{lk})~.
\eeqa
For simplicity the effect of all the Yukawa couplings but
$g_t=m_t/(\sqrt{2}m_W\sin\beta)$ has been neglected. 
Analogous to the previous cases, the variables 
$x_{ij}$ denote ratios of squared masses
(e.g. $x_{ik}=M^2_{\chi_i}/M^2_{\tu_k}$) and the
functions $k(x,y)$ and $j(x,y)$ \cite{BRS} can be found in the appendix.

\FIGURE{
       \epsfig{file=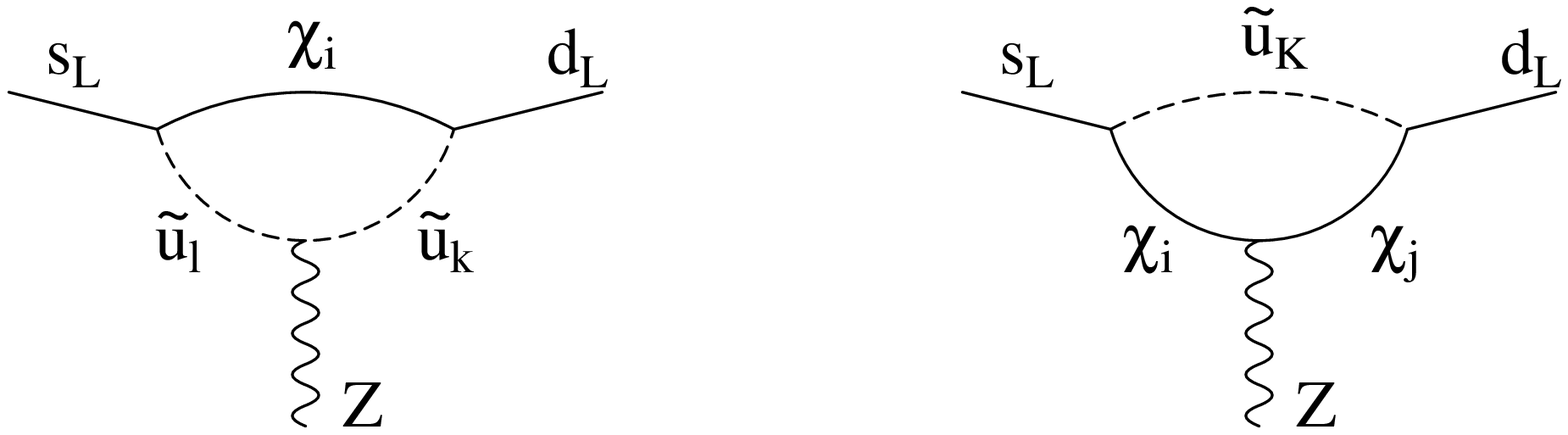,width=12cm}
    \caption{Chargino--up--squark penguin diagrams contributing 
to the $Z{\bar s}d$ effective vertex (diagrams involving self--energy 
corrections to the external legs are not explicitly shown).}
    \protect\label{fig1}
}

The product of $A^d_{jl}$ and ${\bar A}^s_{ik}$ in (\ref{eq:Wchi})
generate four independent terms, proportional to 
$g^2_t\lambda_t$, $g_t V_{td}$, $g_t V_{ts}^*$ and $1$, respectively, 
which correspond to the so--called RR, LR, RL and LL contributions
in the notation of \cite{NW,BRS}. We proceed to analyze these 
terms separately. 

\begin{enumerate}
\item{}
The RR contribution is generated by a Yukawa--type 
interaction in both quark--squark--chargino vertices of
Fig.~\ref{fig1}. This term is 
the only one which survives in the limit of diagonal $\cM^2_U$, 
i.e. to the lowest order in the perturbative 
expansion of ${\hat H}$ around unity. In this limit 
$W_{ds}^{\chi}\vert_{RR}$ is given by 
\beqa
W_{ds}^{\chi}\vert^0_{RR} &=& {1\over 8} g^2_t\lambda_t 
{\hat V}\da_{2j} \Big[ {\hat V}_{j1} {\hat V}_{1i}\da  k(x_{i
  t_R},x_{j t_R}) 
\nonumber \\
&& \qquad\qquad  - 2  {\hat U}_{i1} {\hat U}\da_{1j}  \sqrt{x_{i t_R} x_{j t_R}} 
j(x_{i t_R },x_{j t_R}) \Big] {\hat V}_{i2}~.
\label{eq:WRR}
\eeqa
As anticipated in the introduction, the $SU(2)_L$ breaking 
of the $Z{\bar s}d$ vertex requires at least two 
mass--mixing terms, either from the squark
sector or from the chargino sector. In (\ref{eq:WRR})
the absence of the former mechanism implies a double 
wino--higgsino mixing, as can be easily checked 
by the mismatch of ${\hat V}$ and ${\hat U}$ indices.
Thus $W_{ds}^{\chi}\vert^0_{RR}$ is parametrically 
suppressed by $\cO(m^2_W/M^2_2)$ and aligned in phase
with respect to $W_{ds}^{SM}$. To get a feeling of the
numerical factors, note that $k(1,1)=3/2$ and $j(1,1)=1/2$. 
We then conclude that this contribution cannot provide
a sizable effect, particularly in the limit of 
a heavy supersymmetry--breaking scale.

Considering higher orders in the perturbative expansion of 
${\hat H}$, one can easily check that there is no 
contribution to $W_{ds}^{\chi}\vert_{RR}$ at the first order. 
At the second order it is possible to generate a non--vanishing
contribution and also to avoid the wino--higgsino mixing.
However, the unavoidable factor $\lambda_t$ makes  
$W_{ds}^{\chi}\vert_{RR}$ always not particularly interesting 
with respect to $W_{ds}^{SM}$.

\item{}
The LR and RL terms are originated by a Yukawa--type interaction
in one of the two quark--squark--chargino vertices of
Fig.~\ref{fig1} and a gauge--type interaction in the other.
As can be easily understood, this  implies that
$W_{ds}^{\chi}\vert_{LR}$ and  $W_{ds}^{\chi}\vert_{RL}$
are at least of first order in both wino--higgsino and 
$\tq_L-\tq_R$ mixing. Performing explicitly the expansion
of ${\hat H}$ up to the first order, as discussed in the appendix,
we find
\beqa
&& W_{ds}^{\chi}\vert^1_{LR}\, =\,  - {1\over 8} g_t V_{td} 
~\frac{(M^2_U)_{s_L t_R}}{M^2_{\tu_L}}~
{\hat V}\da_{2j} \Big[ {\hat V}_{j1} {\hat V}\da_{1i} 
k(x_{i u_L}, x_{j u_L}, x_{t_R u_L}) 
\nonumber\\&&\quad -\delta_{ij}  k(x_{i u_L}, x_{t_R u_L}, 1) 
- 2  {\hat U}_{i1} {\hat U}\da_{1j}  \sqrt{x_{i u_L} x_{j u_L}} 
j(x_{i u_L},x_{j u_L}, x_{t_R u_L}) 
\Big] {\hat V}_{i1}~,\quad\quad
\label{eq:WLR}
\eeqa where $M_{\tu_L}$ indicates the average mass of the approximate
left--handed up squarks.  The $W_{ds}^{\chi}\vert^1_{RL}$ term can be
obtained from (\ref{eq:WLR}) with the substitution $V_{td} (M^2_U)_{s_L
t_R}$ $\to$ $V_{ts}^* (M^2_U)_{t_R d_L}$ and with the exchange $1
\leftrightarrow 2$ in the indices of the ${\hat V}$ matrices outside the
square brackets.

The presence of a single CKM matrix element in (\ref{eq:WLR}) leads to
potentially large effects: the missing factor $V_{ts}^*$ is replaced by
$\deu_{ts}^*$, where \beq \deu_{ab}=(M^2_U)_{a_R b_L}/M^2_{\tu_L}~, \eeq
and the ratio $\deu_{t s}/V_{ts}$ can be larger than one
\cite{MPR}. However, this enhancement is partially compensated by the
$\cO(m_W/M_2)$ suppression induced by the wino--higgsino mixing and the
total effect is not very large.  Indeed the phenomenological bounds on
$\deu_{t s}$, dictated mainly by $b\to s \gamma$, become weaker for large
supersymmetric masses, when the wino--higgsino suppression gets stronger.
As a result, $W_{ds}^{\chi}\vert^1_{LR}$ can be at most as large as
$W^{SM}_{ds}$ \cite{BRS}.  Similar comments apply also to
$W_{ds}^{\chi}\vert^1_{RL}$ (see the next section for a more detailed
discussion about limits on $\deu_{t s}$ and $\deu_{t d}$).

It is interesting to note how the LR and RL terms, which arise only at
first order in the expansion of ${\hat H}$, are potentially larger and not
aligned in phase with respect to the lowest order contribution, provided by
$W_{ds}^{\chi}\vert^0_{RR}$. This is clearly a consequence of the
disappearance of one of the two CKM factors. For this reason, it is natural
to expect that terms arising at the second order in the mass--insertion
expansion and without any CKM suppression could be even bigger.

\item{}
The LL term is originated by a double gauge--type interaction
in the quark--squark--chargino vertices of Fig.~\ref{fig1}.
Similarly to the RR case, this implies a second--order
mixing either in the chargino sector or in the $\tq_L-\tq_R$ sector.
However, contrary to the RR case, there is no 
contribution to the leading--order in the expansion of ${\hat H}$. The first 
term in this expansion arises to the first order and 
is given by 
\beqa
W_{ds}^{\chi}\vert^1_{LL} & =&  - {1\over 8}
~\frac{(M^2_U)_{s_L d_L}}{M^2_{\tu_L}}~
{\hat V}\da_{1j} \Big[ {\hat V}_{j2} {\hat V}_{2i}\da 
k(x_{i u_L}, x_{j u_L}, 1) 
\nonumber\\ &&\qquad\qquad\qquad
- 2  {\hat U}_{i2} {\hat U}\da_{2j}  \sqrt{x_{i u_L} x_{j u_L}} 
j(x_{i u_L},x_{j u_L},1) 
\Big] {\hat V}_{i1}~.\quad\quad
\label{eq:WLL}
\eeqa
As can be noted, this term involves a double wino--higgsino 
mixing, which provides the necessary $SU(2)_L$ breaking,
and a first--order mixing among left--handed squarks. 
Thus, even if apparently enhanced by the absence of any 
CKM factor, $W_{ds}^{\chi}\vert^1_{LL}$ is strongly 
suppressed in the  limit
of heavy super\-sym\-me\-try--breaking scale. Moreover, 
the $SU(2)_L$ invariance of the soft--breaking terms 
relates $(M^2_U)_{s_L d_L}$
to $(M^2_D)_{s_L d_L}$ \cite{MPR}, which is strongly constrained 
by $\Ko-\Kob$ mixing. As a result, 
$W_{ds}^{\chi}\vert^1_{LL}$ turns out to be 
always smaller than $W_{ds}^{SM}$  \cite{BRS}.

A different scenario occurs if we consider 
the contribution to  $W_{ds}^{\chi}\vert_{LL}$
which survives in the absence of wino--higgsino 
mixing. In this case one has to go at least
to the second order in the expansion of ${\hat H}$,
and only terms with a double LR mixing survive. 
The lowest--order result in this limit is simply given by
\beqa
W_{ds}^{\chi}\vert^2_{LL} &=&   {1\over 8}
~\frac{(M^2_U)_{s_L q_R}(M^2_U)_{q_R d_L}}{m^4_{\chi_1}}~
l(x_{u_L 1}, x_{u_L 1}, x_{q_R 1} ) \nonumber\\
&=& {1\over 8}
~\deu_{q s}^*~\deu_{q d} ~ x_{u_L 1}^2~
l(x_{u_L 1}, x_{u_L 1}, x_{q_R 1} )~, 
\label{eq:WLL2}
\eeqa
where the function $l(x,y,z)$,  normalized to 
$l(1,1,1)=-1/12$, is reported in the appendix.
Contrary to the cases of $W_{ds}^{\chi}\vert^0_{RR}$,
$W_{ds}^{\chi}\vert^1_{LR}$ and $W_{ds}^{\chi}\vert^1_{LL}$  
discussed previously,
there is no explicit 
suppression in $W_{ds}^{\chi}\vert^2_{LL}$
in the limit of heavy superpartners. 
Actually the upper bounds on the factor $\deu_{q s}^*\deu_{q d}$ go to zero
in this limit, as we will see in the next section.
However, for $q=t$ there is room enough to produce sizable effects
(in close analogy to the $\lambda_t$ factor in the SM case)
even for $M_S\sim 1$ TeV.
In this case $W_{ds}^{\chi}\vert^2_{LL}$ could be substantially 
larger than $W_{ds}^{SM}$ providing sizable 
enhancements to rare kaon decay rates.
\end{enumerate}

\noindent
Using the effective Lagrangian (\ref{eq:Lzsd})
we can easily calculate the effects of the $Z{\bar s d}$ 
penguins discussed above in various processes. 
In the case of $K\to\pi\nu{\bar \nu}$ decays 
we find that the contribution generated by (\ref{eq:Lzsd})
to the $X$ function, defined by 
\beq
\cH_{eff}\, =\, 
  {G_F\over \sqrt{2}}~ \frac{\alpha}{2 \pi \sin^2\Theta_W }~ 
 \lambda_t X ~ {\bar s} \gamma^\mu (1-\gamma_5) d ~{\bar \nu_l} \gamma^\mu 
(1-\gamma_5) \nu_l\, + \, \mbox{h.c.},
\label{eq:Hnn}
\eeq
is given by 
\beq
X_{Z{\bar s}d}=W_{ds}/\lambda_t~.
\label{eq:X_Zds}
\eeq
Comparing 
our results in (\ref{eq:Higgs}), (\ref{eq:WRR}),  (\ref{eq:WLR}) and
(\ref{eq:WLL})   
with those reported in the appendix of \cite{BRS} we find a 
perfect agreement but for an overall factor $-1/2$
which is a mere misprint.\footnote{~This misprint does not 
affect the numerical results of \protect\cite{BRS}. 
We thank L. Silvestrini for clarifying this point.} 

In principle, in the case of  $K\to\pi\nu{\bar \nu}$ decays
also supersymmetric box diagrams could provide sizable effects, 
as it happens for instance in the SM case \cite{BB}. 
However, the contribution of chargino--up--squark box diagrams 
to $X$ turns out to be always suppressed by a factor 
$m^2_W/M^2_\tq $, besides possible wino--higgsino 
mixing.\footnote{~The contribution of the charged--Higgs
box diagrams is clearly negligible because of the small
lepton Yukawa couplings.} In a generic expansion in powers of 
off--diagonal mass terms, denoted by $\epsilon$,
the box contribution to $X$ 
starts at $\cO(\epsilon^3)$, whereas the penguin one at 
$\cO(\epsilon^2)$.
Thus, in general we agree with the statement of Nir and Worah 
\cite{NW} that penguin contributions provide the dominant effect.
Only in the case of the terms 
proportional to $(M^2_U)_{s_L d_L}$, when the penguin 
contribution is suppressed and starts at $\cO(\epsilon^3)$,
the corresponding box term turns out to be competing \cite{BRS}.
However, as long as we are interested only in possible large
effects this is not a relevant case. 

Similar arguments apply also to other processes where the 
effective $Z{\bar s}d$ vertex can contribute, like
$K\to\ell^+\ell^-$ and 
$K\to\pi\ell^+\ell^-$ decays. Hence, in a minimal 
supersymmetric extension of the SM with generic flavor sector,
and particularly in the limit of 
a heavy supersymmetry--breaking scale, we consider it
a good approximation
to encode the dominant non--SM effects to these processes
via the Lagrangian (\ref{eq:Lzsd}). A similar approach
was considered by Nir and Silverman in a different context \cite{NS}:
the coupling $W_{ds}$ in (\ref{eq:Lzsd})
is related to the $U_{ds}$ of  Nir and Silverman  by
\beq
U_{ds} =  \frac{\alpha}{\pi \sin^2\Theta_W} W_{ds}~.
\label{eq:Uds}
\eeq

\section{Bounds on the $\deu_{ij}$ couplings.}
In the previous section we have argued that the (LR)$^2$ term that
appears at second order in the mass--insertion expansion, may give the
largest enhancement to the $Z{\bar s}d$ effective vertex
with respect to the SM contribution. In the present
section we will analyze in detail the bounds we can put on this term,
considering both phenomenological information, and purely theoretical
constraints.

\subsection{Vacuum--stability bounds.}
Before analyzing the bounds coming from phenomenology, we discuss an
interesting result obtained by Casas and Dimopoulos \cite{CaDimo}, who have
shown that bounds on the off--diagonal LR entries of the 
squark mass matrices can be derived also from the requirement that the 
standard vacuum of the theory be stable. 
In particular they require that there are no charge and color
breaking minima (CCB bounds), nor directions in which the potential is
unbounded from below (UFB bounds). Obviously, these bounds have to be
satisfied by any model (and interestingly enough, are generally {\em not}
satisfied).
The only way to avoid or at least to soften these constraints is to assume 
that we live in a sufficiently long--lived metastable vacuum. However,
to be more conservative, we will not take into account this possibility. 
The consequence of the stability bounds for the matrix elements of 
our interest can be stated in a very simple manner:
\begin{equation}
\left|\deu_{ij}\right| \leq m_{u_k} \frac{ \sqrt{2 M^2_\tu+ M^2_\tl }}{ M^2_\tu }
\;\;\;\;\;  \left(k=\mbox{max}(i,j) \right)~,
\label{MI_bounds_1}
\end{equation}
where $M^2_\tu$ and $M^2_\tl$ denotes the average masses of 
up--squarks and sleptons, whereas $m_{u_k}$ indicates the 
mass of the $u_k$ quark. The $m_{u_k}$ factor provides
a very stringent suppression unless one of the two
generation indices ($i$ and $j$) is equal to 3. 
For this reason it is a good approximation to replace 
the sum $\sum_q \deu_{q s}^*~\deu_{q d}$  in (\ref{eq:WLL2})
with the product  $\deu_{t s}^*~\deu_{t d}$.
In this case the bound (\ref{MI_bounds_1})
can be roughly expressed in the following form
\begin{equation}
\left| \deu_{ts}^* \deu_{td} \right| \leq  {3 m^2_{t} \over M^2_S}~,
\label{MI_bounds_2}
\end{equation}
where with $M_S$ we have indicated a typical supersymmetric scale. 
Actually the bound (\ref{MI_bounds_1}) corresponds to the UFB constraint,
but as long as we consider almost degenerate supersymmetric 
particles CCB and UFB bounds are essentially equivalent \cite{CaDimo}. 
 
At this point it is useful to make a first estimate of
the possible enhancement induced by the (LR)$^2$ mass insertion
in the $Z{\bar s}d$ vertex. Comparing (\ref{eq:SM}) and 
(\ref{eq:WLL2}), in the limit $x_{tW} \gg 1$ and assuming 
almost degenerate supersymmetric particles, leads to
\beq
\left| \frac{ W_{ds}^{\chi}\vert^2_{LL} }{ W_{ds}^{SM} } \right|
\simeq
\left| \frac{\deu_{t s}^*~\deu_{t d} }{12~x_{tW}~\lambda_t} \right|
\lsim 20 \times \left( {500\mbox{GeV} \over M_S} \right)^2~,
\eeq
where the last inequality has been obtained imposing 
the bound (\ref{MI_bounds_2}). As can be noticed,
though stringent the model--independent
constraint leaves enough room for a large enhancement, 
even for $M_S$ as large as 1 TeV.

\subsection{Box $\Delta S=2$.}
A term with two LR mass insertions appears in the box diagram
(containing charginos and squarks) contributing to $\Ko-\Kob$ mixing.
In this case, however, this term appears only at a subleading
level. The complete expression for the contribution of the box diagram to
the effective Hamiltonian for $\Delta S =2$ is
\begin{equation}
{\cal H}^{\mbox{\tiny{eff}}}_{\Delta S =2} = {G_F \over \sqrt{2}} { \alpha
  \over \pi \sin^2\Theta_w }~A^d_{ik}\bar{A}^s_{jk} A^d_{jl}\bar{A}^s_{il}
~\frac{m^2_W}{M^2_{\tq_k}}~ k(x_{ik},x_{jk},x_{lk}) (\bar{s}_L \gamma^\mu d_L)  
(\bar{s}_L \gamma_\mu d_L)~.
\end{equation}
We may now expand the mass matrices around their diagonal part to the
desired order. Considering only terms without 
chargino mixing, we get

\begin{eqnarray}
  && A^d_{ik}\bar{A}^s_{jk} A^d_{jl}\bar{A}^s_{il}~ k(x_{ik},x_{jk},x_{lk}) 
         \, =  \nonumber\\
  &&\qquad -\frac{1}{20} \left[(\deull_{sd})^2-\frac{2}{3} \deull_{sd} (
  \deu_{ts}^* 
   \deu_{td} ) +  \frac{1}{7} (\deu_{ts}^* \deu_{td} )^2 + \ldots \right]
   \nonumber \\    
   &&\qquad - {g_t^2 \lambda_t \over 10} \left( 
   \deu_{ts}^* \deu_{td} + \ldots \right) + \cO(\lambda_t^2)~.
\label{BoxDS2}
\end{eqnarray}

To obtain this result we have not only applied the formulae for the
perturbative diagonalization of the mass matrices (that we give in
appendix), but have also taken the limit where all superpartners have 
approximately 
the same mass ($x_{ki}=1$ for all $k$'s and $i$'s).  If we now use the
experimental information on $\Delta m_K = 3.5 \times 10^{-12}$, and require
that the contribution of the term with two LR insertions in
(\ref{BoxDS2})\footnote{~Evaluating this with the approximation $\langle
\Ko | (\bar{s}_L \gamma^\mu d_L) (\bar{s}_L \gamma_\mu d_L)| \Ko \rangle
=~1/3~m_K~f_K^2$~.} does not exceed the experimental value, we get
\begin{equation}
\sqrt{ \mbox{Re} \left[ \left(\deu_{ts}^* \deu_{td} \right)^2 \right]
  } \leq 0.16  \times \left( {M_S \over 500 \mbox{GeV}} \right) \; \; . 
\label{eq:box25}
\end{equation}
We remark that this limit is derived using the quadratic term in (\ref{BoxDS2}), 
as the linear one is multiplied by $\lambda_t$ which suppresses its contribution
strongly. Similarly, we have not considered the bounds that could be obtained 
on the single $\deu_{ts}$ and $\deu_{td}$
couplings, which always appear suppressed both by CKM
factors and chargino mixing.
Of course this limit is rather generous, as one would expect the
first two terms in (\ref{BoxDS2}) to be responsible for the main part
of the effect. On the other hand, until we will be able to get some
independent information on the first two terms in the expansion (and on
their signs too) this is the best we can get from this quantity.

If we now look at the imaginary part of the same matrix element, and
consider the experimental information on Re$(\epsilon)$, we can get 
a bound on the imaginary part of the (LR)$^2$ term squared:
\begin{equation}
\sqrt{ \mbox{Im} \left[ \left(\deu_{ts}^* \deu_{td} \right)^2 \right]
  } \leq 0.015  \times \left( {M_S \over 500 \mbox{GeV}} \right) \; \; . 
\end{equation}

\subsection{Limits from $B$ and $D$ physics.}
Buras, Romanino and Silvestrini \cite{BRS} have analyzed the bounds on
various mass insertions coming from $B$--meson phenomenology.
From the chargino contribution to $B_d-\bar{B}_d$ mixing they get
\begin{equation}
\left| \deu_{dt} \right| \leq 0.1 \times \left( {M_S \over 500
    \mbox{GeV}} \right)~.
\end{equation}
A bound on the other matrix element of our interest
was earlier obtained by  Misiak, Pokorski and Rosiek 
analyzing the chargino contribution to $b \rightarrow s \gamma$ \cite{MPR}:
\begin{equation}
\left| \deu_{st} \right|  \leq 3  \times \left( {M_S \over 500
    \mbox{GeV}} \right)^2~.
\end{equation}

In principle a limit on $\mbox{Re}[( \deu_{ts}^* \deu_{td} )^2 ]$,
similar to the one in (\ref{eq:box25}), could be 
obtained from the analysis of the gluino--up--squark box diagram  
contributing to $D^0-{\bar D^0}$
mixing. Note, however, that this bound is very different from those
discussed above since it  can be made arbitrarily small in the limit of a
heavy gluino mass.  
Assuming gluino and chargino approximately degenerate, the
constraint obtained by $D^0-{\bar D^0}$ mixing 
is essentially  equivalent to the $\Ko-\Kob$ one. 
Indeed the $(g_{\rm strong}/g)^4$ enhancement of 
the gluino box diagram with respect to the chargino one 
is almost completely compensated by the less stringent experimental 
constraint on $\Delta m_D$ with respect to $\Delta m_K$.\footnote{~We are  
grateful to M. Worah for a clarifying discussion about this point.~}

\subsection{Bounds on the $Z{\bar s}d$ vertex.}
As anticipated in the previous section, the $Z{\bar s}d$ effective vertex
contributes to various rare kaon transitions. Some of them have been observed,
whereas stringent experimental limits exist on the others: we can therefore 
use these informations to derive bounds on the (LR)$^2$ term which we are
now analyzing.  
These bounds are best expressed in terms of the coupling
$W_{ds}$ introduced in (\ref{eq:Lzsd}).  A similar analysis has been
already made by Grossman and Nir \cite{GroNir}, using exactly the same 
language of an effective $Z{\bar s}d$ coupling (but using the $U_{ds}$ of
\cite{NS}). 
Following and partially updating (and correcting) their results, we find
\begin{enumerate}
\item
from the process $K_L \rightarrow \mu^+ \mu^-$ \cite{KLmumu}:
\begin{equation}
\left| \mbox{Re} (W_{ds}) \right| \leq 2.2 \times 10^{-3}~;
\label{eq:29}
\end{equation}
\item
from\footnote{~Note that the corresponding bound in
  \protect\cite{GroNir} was larger due to missing factor six
  in their Eq.~(13).}
$B(K^+ \rightarrow \pi^+ \nu \bar\nu) < 2 \times 10^{-9}$ \cite{BNL787}:
\begin{equation}
\left| W_{ds} \right| \leq 3.6 \times 10^{-3}~;
\end{equation}
\item
from the measurement of $\epsilon$:
\begin{equation}
\left| \mbox{Re} (W_{ds}) \mbox{Im} (W_{ds}) \right| \leq 1.1 \times
10^{-5}~.
\label{eq:31}
\end{equation}
\end{enumerate}
In principle similar bounds could be obtained from $\epsilon'/\epsilon$
and $B(K_L \rightarrow \pi^0 e^+ e^-)$. 
However, in both cases the large theoretical uncertainties 
and the poor experimental information lead to weaker constraints.  
To translate the results (\ref{eq:29}-\ref{eq:31}) 
into bounds for the (LR)$^2$ term of our
interest we use the relation
\[
W_{ds}^{\chi}\vert^2_{LL} =\frac{1}{96} \deu_{ts}^* \deu_{td}~,
\]
derived from (\ref{eq:WLL2}) in the limit of degenerate 
supersymmetric particles. We then obtain
\begin{eqnarray}
\left| \mbox{Re} \left(\deu_{ts}^* \deu_{td} \right) \right| & \leq &
0.21~, \nonumber \\ 
\left| \deu_{ts}^* \deu_{td} \right| & \leq & 0.35~,\nonumber \\
 \left| \mbox{Re} (\deu_{ts}^* \deu_{td})
\mbox{Im} (\deu_{ts}^* \deu_{td}) \right| & \leq & 0.1~.
\end{eqnarray}

\subsection{Summary of the bounds.}
The two limits derived from the analysis of the
$\Delta S=2 $ box diagram can be written as follows:
\begin{eqnarray}
\left| \left[ \mbox{Re} \left(\deu_{ts}^* \deu_{td} \right) \right]^2 
  - \left[ \mbox{Im} \left(\deu_{ts}^* \deu_{td} \right) \right]^2
\right|\! & \leq \! & 2.6 \times 10^{-2} \times \left( {M_S \over 500
    \mbox{GeV}} 
\right)^2 \; ,\qquad \label{eq:Re_box_lim} \\
\left| \mbox{Re} (\deu_{ts}^* \deu_{td})
\mbox{Im} (\deu_{ts}^* \deu_{td}) \right| 
\! & \leq \! & 1.1 \times 10^{-4} 
\times \left( {M_S \over 500 \mbox{GeV}} \right)^2 \; ; \qquad
\label{eq:Im_box_lim}
\end{eqnarray}
those obtained from $B$ physics lead to
\begin{equation}
\left| \deu_{ts}^* \deu_{td} \right| \leq 0.3 \times \left( {M_S
\over 500 \mbox{GeV}} \right)^3~,
\end{equation}
whereas the model--independent one is given by
\begin{equation}
\left| \deu_{ts}^* \deu_{td} \right| \leq 0.3 \times \left( { 500
    \mbox{GeV} \over M_S} \right)^2~.
\label{eq:mib}
\end{equation}
Finally, the `scale--independent' limits
derived from the phenomenological 
analysis of the $Z{\bar s}d$ vertex are
\begin{eqnarray}
\left| \mbox{Re} \left( \deu_{ts}^* \deu_{td} \right) \right| & \leq
& 0.21~,\nonumber \\ 
\left| \deu_{ts}^* \deu_{td} \right| & \leq & 0.35~,
\label{eq:bound2}
\end{eqnarray}
where we have skipped the bound on the product of real and imaginary part,
which is clearly negligible with respect to the one in (\ref{eq:Im_box_lim}).

\FIGURE{       
  \epsfig{file=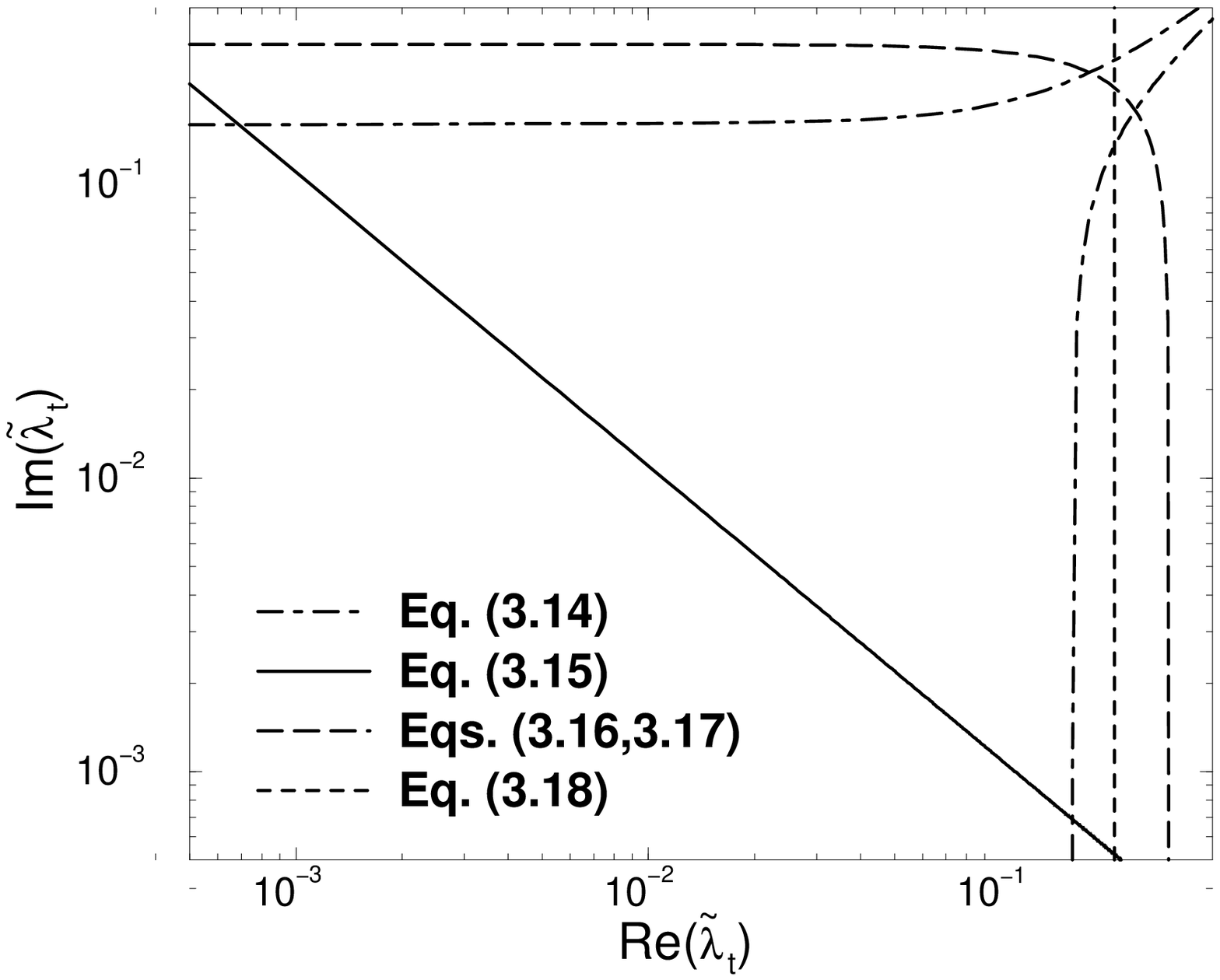,width=12cm}
    \caption{Summary of the bounds on the (LR)$^2$ coupling
${\tilde \lambda}_t$, as defined in Eq.~(\protect\ref{eq:ldef}).}
\protect\label{fig2}
}
A summary of the various bounds is displayed in Fig.~\ref{fig2}, for the
sample value $M_S=500$ GeV. From the figure it is clear that the bound in
(\ref{eq:Im_box_lim}) is by far the most stringent one. This implies that,
if we assume that real and imaginary parts of $\deu_{ts}^* \deu_{td}$ are
of the same order, these are $\cO(10^{-2})$.  On the contrary, if one of
the two is zero the other can be $\cO(10^{-1})$.  The maximum value allowed
for the real part setting the imaginary part to zero, or vice versa, is 0.16
as dictated by the bound 
(\ref{eq:Re_box_lim}).  Playing around with
the $M_S$ dependence of these bounds one can find that the maximum value
allowed for either the imaginary or real part (when the other is set to
zero) can grow up to 0.2 for $M_S=600$ GeV, again a bound dictated by
(\ref{eq:Re_box_lim}).  Above this value of $M_S$ the
model--independent limit on the modulus becomes more stringent.  Notice in
fact that the model--independent limit on $\left| \deu_{ts}^* \deu_{td}
\right|$ and the one coming from $B$--physics have opposite dependence on
the average mass of the superpartners.  So that for $M_S < 500 $ GeV it is
the $B$--physics one which dominates, whereas above 500 GeV the
model--independent one takes over.  

In conclusion, for $M_S\gsim 600$ GeV  we can just consider the two 
bounds in  (\ref{eq:Im_box_lim}) and (\ref{eq:mib}), as
all the others will be automatically satisfied.
If we define 
\beq
{\tilde \lambda}_t= | {\tilde \lambda}_t  | e^{ i {\tilde \theta}_t }
 = \deu_{ts}^* \deu_{td}~,
\label{eq:ldef}
\eeq
the bound we have to satisfy for $M_S \gsim 600$~GeV is
\beq
|{ \tilde \lambda}_t  | < {\rm min} \left[0.2 \times \left( {  
  600  \mbox{GeV} \over M_S } \right)^2, \frac{ 2\times 10^{-2} } {\sqrt{ |
  \sin 2{\tilde   \theta}_t |} } \left( { M_S 
  \over 600  \mbox{GeV} } \right) \right] ~,
\label{eq:lastb}
\eeq
whereas the phase ${\tilde \theta}_t$ is unbounded.

At this point one could argue whether a reasonable low--energy 
supersymmetric model could saturate 
this bound. 
It is beyond the scope of this paper to analyze any model in detail. 
However, we recall that in generic superstring scenarios the so--called 
$A$ terms, responsible for the LR entries of the 
mass matrices, are expected to be ${\cal O}(M_S)$ \cite{string}. 
This would imply $(M^2_{LR})_{ij} \sim {\cal O}(m_W \cdot M_S)$.
Thus  in general it is not unnatural to consider models 
where the bound (\ref{eq:lastb}) is saturated
(see e.g. the discussion at the end of Ref.~\cite{CaDimo}).

\section{Phenomenological consequences of a large $Z{\bar s}d$ effective
  coupling} 

The most clear signature of an enhancement in the $Z{\bar s}d$ effective
vertex could be found in $K \to \pi\nu \bar{\nu}$ decays. Within the SM
these transitions can be described by means of the Hamiltonian
(\ref{eq:Hnn}), with the $X$ function given by \cite{BB} \beq X_{SM} =
X_t\left(x_{tW}\right) + \frac{\lambda^4 \lambda_c }{ \lambda_t} P_c~,
\label{eq:X_SM}
\eeq
where $X_t(x_{tW}) \simeq 1.5$ 
is generated by the dominant top--quark contribution (summing penguin and
box diagrams) and $P_c=0.40\pm0.06$ is due to the charm loops
(as usual $\lambda$ denotes the Cabibbo angle and 
$\lambda_c = V_{cs}^*V_{cd}$, 
thus $|\lambda^4 \lambda_c / \lambda_t|\sim \cO(1)$).
The branching ratios of $K^+$ and $K_L$ modes can be 
expressed in terms of the  $X$ function as 
\beqa
BR(K^+ \to \pi^+ \nu \bar{\nu}) &=& \kappa_+ \left| \frac{ \lambda_t X
    }{\lambda^5}  \right|^2~, \\
BR(K_L \to \pi^0 \nu \bar{\nu}) &=& \kappa_L \left( \mbox{Im} \frac{
    \lambda_t X }{\lambda^5}  \right)^2~,
\eeqa
where $\kappa_+=4.11\times 10^{-11}$ and  $\kappa_L=1.80\times 10^{-10}$
\cite{BB2}. 
For a numerical estimate we recall that 
$\lambda= 0.22$, $|\lambda_t| \simeq 3\times 10^{-4}$
and $\mbox{Im}\lambda_t \simeq  |\lambda_t|/3$ \cite{BFle}.

In extensions of the SM where the main new--physics
effects can be encoded via the effective coupling 
$W_{ds}$, we should add to $X_{SM}$ the $X_{Z\bar{s}d}$
function defined in (\ref{eq:X_Zds}). Thus if we add to the SM 
contribution the dominant supersymmetric effect, provided by 
the (LR)$^2$ terms, we find
\beqa
X_{tot} &=&  \frac{1}{8} \frac{{\tilde \lambda}_t }{ \lambda_t}  
~ x_{u_L 1}^2~l(x_{u_L 1}, x_{u_L 1}, x_{q_R 1} ) + 
X_t\left(x_{tW}\right) + \frac{\lambda^4 \lambda_c }{ \lambda_t} P_c
\nonumber \\
&\simeq&  \frac{1}{96} \frac{{\tilde \lambda}_t }{ \lambda_t} +
X_t\left(x_{tW}\right) + \frac{\lambda^4 \lambda_c }{ \lambda_t} P_c~,
\label{eq:X_tot}
\eeqa
where ${\tilde \lambda}_t$ has been defined in (\ref{eq:ldef})
and the second line of (\ref{eq:X_tot}) 
is obtained in the limit of almost degenerate superpartners.
Given the constraints on $|{\tilde \lambda}_t|$ reported
in (\ref{eq:lastb}) it is clear that large enhancements 
with respect to the SM case are possible. 
In the rate of the charged mode one can gain up to an order 
of magnitude if $M_S$ is around  $600$ GeV, 
where the effect is maximum. In the $K_L$ case a crucial 
role is played by the new phase ${\tilde \theta}_t$: 
if  ${\tilde \theta}_t \sim 90^\circ$
a huge enhancement (up to two order of magnitudes) is possible  
for a wide range of $M_S$. In Table~\ref{tab:BR} we summarize 
the upper bounds on the two modes for $M_S\sim 0.6$ TeV and 
$M_S\sim 1$ TeV. 
\TABULAR{|l||c|c|r|}{ \hline  
 & \multicolumn{2}{c|}{ } & \\ 
\ \  decay mode  & \multicolumn{2}{c|}{maximum SUSY branching ratio} 
    &  SM branching ratio $\quad$\ \\ 
& \multicolumn{2}{c|}{ $M_S \sim 0.6$ TeV $\quad$  $ M_S \sim 1$ TeV}   
     &  \\ \hline\hline
 & & & \\
$K^+\to\pi^+\nu\bar{\nu}$ & \ $\quad 1\times 10^{-9} \quad$ \ &  \ $\quad
4\times 10^{-10}  
     \quad$\  &   $(9.1\pm3.8)\times 10^{-11}$ \protect\cite{BFle} \\ 
 & & & \\
\hline
 & & & \\
$K_L\to\pi^0\nu\bar{\nu}$ & $4\times 10^{-9}$ & $6 \times 10^{-10}\ $ 
    &   $(2.8\pm 1.7)\times 10^{-11}$  \protect\cite{BFle} \\ 
 & & & \\
\hline
 & & & \\
$K_L\to\pi^0 e^+ e^-$    & $6\times 10^{-10}$ & $1 \times 10^{-10} $ 
    &   $\lsim$~few$~\times 10^{-11}$ \protect\cite{BFle,DEIP} \\ 
 & & & \\
\hline}
{\label{tab:BR} Approximate upper bounds for the branching ratios 
of $K\to\pi\nu\bar{\nu}$ and $K_L\to\pi^0 e^+e^-$ decays  within
the low--energy supersymmetric scenario discussed in the text,
compared to the SM expectations.}

Related modes which could allow to detect an enhancement in
the $Z{\bar s}d$ effective amplitude are the  
$K\to\pi\ell^+\ell^-$ decays. In the charged channel
the long--distance process $K^+ \to \pi^+ \gamma^*\to \pi^+ \ell^+\ell^-$
is by far dominant, hiding the contribution of the $Z{\bar s}d$ 
transition. However, the single--photon exchange 
amplitude is forbidden by $CP$ invariance in the 
$K_L \to \pi^0 \ell^+\ell^-$ mode, which is therefore more sensitive 
to short--distance dynamics 
(see e.g. Ref.~\cite{DEIP} for a recent discussion about these decays).
Assuming that both $K_L \to \pi^0 e^+ e^-$ and $K_L \to \pi^0 \nu \bar{\nu}$
transitions are dominated by the $CP$--violating part of the  
$Z{\bar s}d$ effective amplitude, we can easily relate their widths.
Indeed, neglecting the electron mass and the effects of the small
vector coupling of the electrons to the $Z$, leads to
$\Gamma(K_L \to \pi^0 e^+ e^-) = \Gamma(K_L \to \pi^0 \nu \bar{\nu})/6$.
Using this approximate relation we have derived the upper bound 
for $B(K_L \to \pi^0 e^+ e^-)$ reported in the last line of 
Table~\ref{tab:BR}. As it is well known, in addition to the 
direct $CP$--violating transition, the  $K_L \to \pi^0 e^+ e^-$
decay can proceed through indirect $CP$--violation 
($K_L \to K_S \to \pi^0 e^+ e^-$) or via the $CP$--conserving  
two--photon exchange ($K_L \to \pi^0 \gamma\gamma \to \pi^0 e^+ e^-$).
However, both these mechanisms are expected to produce 
corrections to $B(K_L \to \pi^0 e^+ e^-)$ at the level of 
few$\times10^{-11}$ at most \cite{BFle,DEIP}. This ensures that
a detection of $B(K_L \to \pi^0 e^+ e^-)$ above $10^{-10}$ can be 
considered as a clear signature of new--physics.

As can be expected, the upper limits for the supersymmetric branching
ratios shown in Table~\ref{tab:BR} are much larger than those
reported in \cite{SusyKpnn}, where the supersymmetric contributions to
$K\to\pi\nu\bar{\nu}$ have been evaluated essentially without allowing
$({\tilde d},{\tilde s})_L-{\tilde t}_R$ mixing. We stress, however, that
our upper bounds are significantly larger also than those
recently obtained in \cite{BRS}, where the $({\tilde d},{\tilde
s})_L-{\tilde t}_R$ mixing has been evaluated only to first order
within the mass--insertion approximation.

To conclude this section we emphasize that in these numerical results we
did not take into account other possible sources of enhancement.
Indeed we could have obtained even
bigger effects playing with the various supersymmetric mass ratios (that we
have set to 1 just to simplify our results).  In particular, larger effects
are obtained with a wino mass lighter than the average squark
mass. Moreover, we have neglected possible constructive interferences
between the leading (LR)$^2$ terms and the subleading, but still not
negligible, LR terms. Finally, we have neglected possible destructive
interferences between chargino-- and gluino--mediated amplitudes when
evaluating the bounds on the LR couplings induced by $\Ko-\Kob$ mixing:
this effect could easily lead to overcome the stringent 
constraint in Eq.~(\ref{eq:Im_box_lim}).

\section{Conclusions}
In this paper we have analyzed supersymmetric contributions to rare $K$
decays mediated by an effective $Z\bar{s}d$ vertex. We have adopted the
strategy of the so--called mass--insertion approximation, which consists in 
assuming that the squark mass matrices are almost diagonal, and that their
diagonalization can be performed perturbatively.
While recent similar analyses have stopped this approximation to the first
order, we have argued that in the present case
it is necessary to go up to the second order in
this expansion to account for all possible important effects.

This result does not contradict the validity of the mass--insertion
approximation. Rather, we have stressed the fact that there is an interplay
between the squark mass matrices and other mass matrices present in the
theory. The reason why the second--order terms 
in this expansion can be more
important than the first--order ones, is because they do not contain
anymore off--diagonal CKM matrix elements which are known to be suppressed.
In other words, we could say that all mass matrices (both those of the
quarks and of the squarks) in the supersymmetric theory are almost
diagonal, and that for all these matrices we count off--diagonal elements
as of order $\epsilon$.
According to this counting rule, both the SM and the SUSY contributions to
this process are of order $\epsilon^2$, and here we have for the first time 
presented a complete result of SUSY effects at order $\epsilon^2$.

Moreover, for reasons related to the necessary presence of
$SU(2)_L$--breaking effects in the effective $Z\bar{s}d$ vertex, 
the supersymmetric contributions generated at the 
first order in the mass--insertion always appear suppressed by 
off--diagonal elements of the {\em chargino} mass matrix. These 
vanish as $\cO(m_W/M_S)$ in the limit of a large supersymmetric 
scale, $M_S$, thus providing an additional damping factor
which can be avoided only at the second order in the 
expansion of the squark mass matrices. 
This suppression as well as the CKM one can only be avoided considering a
double LR mixing in the up--squark sector.

We have performed a numerical analysis of the present
bounds on the off--diagonal LR elements of the up--squark mass matrix
relevant to the effective $Z\bar{s}d$ vertex.  
As a result, we have found that to our present knowledge the term
which had been neglected so far (i.e. the one generated at second order 
in the mass--insertion approximation) is the most dangerous one, and could lead
to very large enhancements in rare kaon decay rates.  
We have shown that the
$K^+ \rightarrow \pi^+ \nu \bar\nu$ rate could be enhanced up to one order
of magnitude with respect to the SM prediction, whereas the neutral decay
mode $K_L \rightarrow \pi^0 \nu \bar\nu$ could be enhanced by up to two
orders of magnitude. The same two orders of magnitude enhancement could be
produced also in the decay $K_L \rightarrow \pi^0 e^+ e^-$.
Finally, we have also briefly discussed why the supersymmetric box 
contributions to these decays can be neglected as long as
we are interested in potentially large effects.

Our results show that the current experimental efforts in the search for
these rare decays are very much welcome and could give us valuable
information on the flavor structure of the soft--breaking terms 
of a generic supersymmetric extension of the SM. Interestingly, we
will not have to wait too long before experiments will reach the  
sensitivity necessary to observe, or at least to constrain,
these supersymmetric effects. Indeed a preliminary 
evidence of the  $K^+ \rightarrow \pi^+ \nu \bar\nu$
decay has been recently obtained \cite{BNL787} and the BNL--E787 Collaboration
is  already analyzing  new data on this mode. A sensitivity on
$B(K_L \rightarrow \pi^0  e^+ e^+)$ at the level of $10^{-10}$ is 
expected in few years by the KTeV experiment at Fermilab \cite{KTeV}. 
Finally, concerning the challenging $K_L \rightarrow \pi^0 \nu \bar\nu$
channel, while waiting for the dedicated experiments aiming to reach a 
sensitivity of $10^{-12}$ \cite{KLNN}, even a non--dedicated experiment 
like KLOE \cite{KLOE} has a chance to give new and valuable 
information on possible extensions of the SM, since it can reach a
sensitivity of 10$^{-9}$ \cite{BCI}.

\section*{Acknowledgments}
We thank S.~Bellucci for participation in the early stage of
this work. We are grateful also to N. Arkani-Hamed, F.~Bossi and M.~Worah 
for interesting discussions. G.I. acknowledges the hospitality 
of the Theory Group at SLAC, where part of this work has been done. 
This project is partially supported by the EEC-TMR Program, Contract N.
CT98-0169.

\newpage
\section*{Appendix}
\newcounter{zahler}
\renewcommand{\theequation}{\Alph{zahler}.\arabic{equation}}
\setcounter{zahler}{1}
\setcounter{equation}{0}

\subsection*{Expansion of the mass matrices around the diagonal.}
Here we report the formulae needed to make the expansion around the
diagonal of the mass matrices up to second order, i.e. including two mass
insertions. Given an $n\times n$ Hermitian matrix $M$, we can
decompose it in the form
\beq
M=M^0+M^1~,
\eeq
where $M^0=\mbox{diag}(m^0_1,\ldots , m^0_n)$ and $M^1$
has no elements on the diagonal. $M$ can be diagonalized by a unitary
matrix $X$, such that $X M X^\dagger = \mbox{diag}(m_1,\ldots , m_n)$.
Then, if $f$ is an arbitrary function, we have
\beqa
X^\dagger_{ik}f(m_k) X_{k j} &=& \delta_{ij} f(m^0_i) + M^1_{ij}
f(m^0_i,m^0_j) \nonumber\\
&& + M^1_{ik} M^1_{kj} f(m^0_i,m^0_j,m^0_k) + O\left( ({M^1})^3
\right)~, 
\eeqa
where we have adopted the notation of Buras, Romanino and Silvestrini
\cite{BRS} to define an $n$--argument function from an $n-1$--argument one:
\begin{equation}
\label{recur}
f(x,y,z_1,\ldots,z_{n-2})=
\frac{f(x,z_1,\ldots,z_{n-2})-f(y,z_1,\ldots,z_{n-2})}{x-y}~.
\end{equation}

\subsection*{Loop functions.}
The loop functions appearing in the 
top--quark penguin diagrams discussed in section~\ref{sect:Zsd}
are given by 
\beqa
C(x)&=&{x\over 8}~\left( \frac{x-6}{x-1} +\frac{3x+2}{(x-1)^2}\log x
\right)~,\\
H(x)&=&{x^2\over 8}~\left( -\frac{\log
x}{(x-1)^2}+\frac{1}{x-1}\right)~.
\eeqa
The multi--variables functions $k(x_1,\ldots,x_n)$, 
$j(x_1,\ldots,x_n)$ and $l(x_1,\ldots,x_n)$, occurring in
chargino--squark diagrams, are defined according to the recursive formula
given in (\ref{recur}).
The explicit expression of the single--variable functions are
\beqa
j(x)&=&\frac{x\log x}{x-1}~,\qquad\qquad k(x)\, =\, x~j(x)~,\\
l(x)&=& k\left({1\over x}~,~{1\over x}\right)
      -{2\over x}j\left({1\over x}~,~{1\over x}\right)
      - k\left({1\over x}~,~{x_{u_L1}\over x}\right)
      - k\left({1\over x_{u_L1}}~,~{x \over x_{u_L 1}}\right)~.\qquad
\eeqa

\end{document}